\documentclass[referee,10pt,a4]{aa} 
\usepackage{graphicx}

\begin{document}

\title{ Dynamical condition of neutral hydrogen envelopes of dwarf
galaxies and their possible morphological evolution }

\author{Yuka Y. Tajiri 
          \and
        Hideyuki Kamaya 
          }

 \offprints{Y.Y. Tajiri}

\institute{Department of Astronomy, Faculty of Science, Kyoto University,
Sakyo-ku, Kyoto 606-8502, Japan\\
             \email{tajiri, kamaya@kusastro.kyoto-u.ac.jp}
             }

   \date{Received September 3rd., 2001; accepted ****}

\abstract{ We investigate the star-formation history of gas-rich dwarf
galaxies, taking account of the dynamical evolution of their neutral
hydrogen (H{\sc i}) envelope.  Gas-rich dwarfs are classified into
blue compact dwarfs (BCDs) and dwarf irregulars (dIrrs). In this
paper, their H{\sc i} envelope is clearly shown not to be blown away
by their stellar feedback. This is concluded since the observed
star-formation rate (SFR) of gas-rich dwarfs is generally smaller than
a critical SFR, $\psi_{\rm crit}$, at which stellar feedback
accelerates the H{\sc i} envelope to the escape velocity.  From this
standpoint and the chemical property of sample BCDs, we suggest two
possibilities; (1) The H{\sc i} gas in the envelope of BCDs is
consumed to fuel their star-formation; and (2) BCDs have a similar
star-formation history. We also discuss morphological evolution among
dwarf galaxies.  As long as gas-rich dwarfs are isolated, it is
difficult for them to evolve into dwarf ellipticals (dEs).  When the
H{\sc i} envelope in gas-rich dwarfs is consumed in subsequent
star-formation, a morphological exchange between BCDs and dIrrs is
still expected, consistent with previous studies.  If the SFR of
gas-rich dwarfs was much higher than $\psi_{\rm crit}$ in the past,
interestingly, an evolutionary scenario from dEs to gas-rich dwarfs is
possible.
}
\titlerunning{Neutral Hydrogen of BCD}
\authorrunning{Y.Y.Tajiri and H.Kamaya}
   \maketitle
%


\section{Introduction}
For a hierarchical cosmic structure formation scenario, the
star-formation history of dwarf galaxies, which are building blocks of
the structure, needs to be known. To determine their star-formation
history, the effect of stellar feedback (e.g., supernova-driven winds
and stellar winds) on the dynamical condition of the interstellar
medium (ISM) of dwarf galaxies is essential, since their gravitational
potential well is rather shallow (Skillman \& Bender 1995 for a
review). Thus, the dynamics of the ISM in dwarfs is a central issue
for their evolution.  In this paper, we study isolated dwarf galaxies
to investigate their evolution, and examine the effect of stellar
feedback on the star-formation rate (SFR) (Larson 1974; Saito 1979;
Dekel \& Silk 1986).

Dwarf galaxies whose star-formation is currently observable have
sufficient amounts of H{\sc i} gas, compared to their dynamical mass
(e.g. Thuan \& Martin 1981). These gas-rich dwarfs are categorized as
dwarf irregular galaxies (dIrrs) and blue compact dwarf galaxies
(BCDs) by their SFR relative to their size. Since their H{\sc i} gas
mostly surrounds their central star-forming regions (e.g. van Zee et
al. 1998), we call it an H{\sc i} envelope in this paper. Although it
is widely believed that the H{\sc i} envelope is gravitationally bound
to host galaxies, the dynamical evolution of the H{\sc i} envelope is
not completely understood.  This is not only because the peaks of the
rotation curves in the H{\sc i} envelopes are not always
satisfactorily determined (e.g. Pustilnik et al. 2001), but also
because there is no confirmation that the H{\sc i} gas in the envelope
is used for a subsequent star-formation (e.g. Thuan 1985). Therefore,
it is useful to investigate the dynamical evolution of the H{\sc i}
envelope from another point of view, that is, in the framework of a
star formation history (see also Legrand et al. 2001).

Indeed, it has been suggested that the H{\sc i} envelope is related to
star-formation activity (e.g. Thuan 1985), which results in the
morphological exchange among dwarf galaxies (e.g. Kormendy
1985). Concerning the morphological evolution, Thuan (1985) suggested
from photometric observations that gas-rich galaxies like BCDs and
dIrrs evolve into gas-less galaxies like dwarf ellipticals (dEs). The
inverse evolution is also theoretically expected: dEs acquire H{\sc i}
gas from the ambient intergalactic medium (IGM), and evolve into dIrrs
and BCDs (Silk et al. 1987).  According to \cite{skt01}, the
morphological exchange between BCDs and dIrrs is also expected, if
H{\sc i} gas in the envelope is channeled into star-forming
regions. Contrary to the above suggestion, it is also shown from the
difference of their surface brightness (Bothun et al. 1986) and their
dynamical (van Zee et al. 2001) properties that gas-rich dwarfs do not
easily evolve into dEs. Thus, it is still premature to judge which
model is correct.  In this paper, we investigate this unclear
star-formation history of gas-rich dwarf galaxies, considering the
dynamical condition of the H{\sc i} envelope which is affected by
stellar feedback.

This paper is organized as follows: In the next section, we construct
a dynamical model of the H{\sc i} envelope from a momentum
conservation law.  The observational data are compared with the model
in section 3.  The implications for the morphological evolution of
dwarfs are given in section 4.  In the last section, we summarize our
conclusions and comment on Legrand et al. (2001), whose conclusions
are very similar to and consistent with those in this paper.

\section{Model description}

In this paper, we try to connect the dynamical condition of the H{\sc
i} envelope and star-formation history of gas-rich dwarf galaxies. For
such a complicated problem, a simple analytical model is very
useful. Indeed, this would allow us to predict the critical limit of a
SFR where gas-rich dwarfs can sustain the H{\sc i} envelope. If a SFR
is larger than the critical one, the H{\sc i} envelope will be blown
away by stellar feedback (e.g. Saito 1979; Dekel \& Silk
1986). Ferrara \& Tolstoy (2000) analyze similar dynamical conditions
of H{\sc i} envelopes in dwarf galaxies with their dynamical mass
being blown-away (all gas is removed), blown-out (partial gas is
removed), or no-mass-loss (no gas is removed) (classification proposed
by De Young \& Heckman 1994).  These hydrodynamical processes are
numerically examined in Mac Low \& Ferrara (1999), and their results
are confirmed by Silich \& Tenorio-Tagle (2001). Silich \&
Tenorio-Tagle (2001) have also shown that the low density halo
suppresses the outflow of gas from the main body of a galaxy.  From
this standpoint, the circulation of the ISM is expected to be like a
chimney (Kunth \& \"Ostlin(2000) for a review). This means that a
closed box approximation is adequate to consider the evolution of
gas-rich dwarfs whose ISM is not blown out or blown away.
To confirm these results with a SFR that can be easily observed
compared to dynamical mass (Pustilnik et al. 2001), we reformulate a
critical value that determines a condition for the H{\sc i}
envelope. Consequently, as shown in the following, we can get
intuitive pictures of the star-formation history and dynamical
properties of the H{\sc i} envelope.  We have newly considered this
useful representation which is similar to that of Legrand et
al. (2001).


Before discussing the critical SFR, we review the stellar feedback
(i.e., supernovae and stellar winds) effect on the H{\sc i}
envelope. Stellar feedback causes turbulence in the ISM, and can make
super-bubbles.  Indeed, turbulence in the H{\sc i} envelope is
observed (van Zee et al. 1998), and super-bubbles are also observed in
some BCDs (e.g. Gil de Paz et al. 1999). In a BCD, the typical size of
a super-bubble is larger than that of its star-forming regions, but is
smaller than that of its H{\sc i} envelope. Martin (1996) thus
suggested that the H{\sc i} envelope must suffer from the stellar
feedback. From the fact that the H{\sc i} envelope is not blown away,
we are able to deduce the allowed strength of the stellar feedback,
the critical SFR, $\psi_{\rm cirt}$ (e.g. Silich \& Tenorio-Tagle
1998).  Legrand et al. (2001) have considered a similar condition and
developed a self-consistent model that treats together both the H{\sc
i} envelope and the low density halo (including dark matter). On this
point, our discussion can provide further insight into the dynamical
evolution of the H{\sc i} envelope.

To determine $\psi_{\rm crit}$, we deduce the following dynamical
condition between the momentum of the H{\sc i} envelope, $p_{\rm
gas}$, and that supplied from the total stellar feedback by
supernovae, $p_{\rm SN}$ (Lamers \& Cassinelli 1999), as being
\begin{eqnarray}
(1 + A_{\rm halo}) p_{\rm gas}>\,p_{\rm SN},
\label{eq:momentum_condition}
\end{eqnarray}
where $A_{\rm halo}$ denotes the contribution from the low density
halo. It can reach at most 1 when a closed box model is satisfied
(Legrand et al. 2001), which is a good assumption, as confirmed in the
next section.  This inequality is called ``the momentum condition''
(e.g. Ferrara \& Tolstoy 2000). Here, we assume that the feedback
energy from stellar winds is comparable to that of supernova
explosions.

The momentum of the H{\sc i} envelope is expressed as $p_{\rm gas}=
M_{\rm gas}\,v_{\rm gas}$, where $M_{\rm gas}$ is the total mass of
the H{\sc i} envelope, and $v_{\rm gas}$ is the out-going velocity of
the H{\sc i} gas.  For simplicity, we adopt the same velocity for the
low density halo to use Eq.(1), which does not alter our main
conclusion since the escape velocity of the low density halo is
generally larger than $v_{\rm gas}$.  In order to estimate $p_{\rm
SN}$, we consider the evolution of a supernova remnant in the Sedov
phase.  As is discussed in Chevalier (1976), the Sedov phase of
expanding supernova remnants starts when the swept-up mass of a shell
of ISM or circumstellar medium is comparable to the ejected mass of a
supernova (e.g. Jones, Smith, \& Straka 1981). Thus, the total
momentum supplied from supernova remnants during a single generation
of star-formation is expressed as $p_{\rm SN}=\eta M_{\rm SN}\,v_{\rm
SN}^2/v_{\rm s}$, where $M_{\rm SN}$ is the total mass of supernova
remnants during the generation, $v_{\rm SN}$ is the typical velocity
of the initial expansion of the remnants, and $v_{\rm s}$ is the final
expanding velocity of the shell. Here, $\eta$ is a coefficient whose
value is adopted as 15/77 from Mac Low \& McCray (1988) and which
denotes net energy reduction owing to the shell formation and the
energy conversion to the thermal energy of the inner coronal gas.
Therefore the momentum condition is re-written as being
\begin{eqnarray}
(1 + A_{\rm halo}) M_{\rm gas}\, v_{\rm gas} &>& \displaystyle
\frac{15}{77} {\frac{M_{\rm SN}\,v_{\rm SN}^2}
{v_{\rm s}}}.
\label{eq:momentum}
\end{eqnarray}

We can relate the right hand side of Eq. (2) to a current SFR.  Assuming
a mass function, $\phi(m)$, we estimate $M_{\rm SN}$ as
\begin{eqnarray}
M_{\rm SN}=\frac{\psi t_{\rm SF}\,
{\displaystyle\int}_{8M_{\odot}}^{100M_{\odot}}m\phi(m)\,dm}
{{\displaystyle\int} _{0.1M_{\odot}}^{100M_{\odot}}m\phi(m)\,dm},
\label{eq:IMF}
\end{eqnarray}
where $\psi$ is a current SFR in the unit of $M_\odot$ yr$^{-1}$, and
$t_{\rm SF}$ is the duration of a single generation of star-formation in
the unit of yr.

The gas mass of the H{\sc i} envelope is known to be dominated by the 
H{\sc i} compared to H$_{2}$ (Barone et al.2000). 
This also means that the
direct contribution of halo gas to the H{\sc i} envelope is small. However,
we take into account halo gas using a coefficient of $(1 + A_{\rm
halo})$ in Eq. (2), because we consider the whole gas system under a
closed box approximation. The out-going velocity $v_{\rm gas}$ is assumed
to be about the escape velocity of the dwarfs with a flat rotation curve,
given as
\begin{eqnarray}
v_{\rm gas}\sim \displaystyle{\sqrt{\frac{2G}{R_{\rm gal}}
{\frac{M_{\rm gas}}{F}}}}
&=& 2.6\times 10^1
\left(\frac{M_{\rm gas}\left(1 + A_{\rm halo} \right) }
           {10^8 M_{\odot}}\right)^{2/7}
\left(1 + A_{\rm halo} \right)^{-2/7}
\left(\frac{F}{0.5}\right)^{-2/7}
{\rm  km\,  sec}^{-1},
\label{eq:v_gas}
\end{eqnarray}
where we adopt the Burkert profile of gas-rich dwarf galaxies (Burkert
1995) using a radius,
$R_{\rm gal}=6.9\times 10^{-4}M_{\rm gas}~ ^{3/7}F^{-3/7}~ {\rm kpc}$.
Here, $G$ is the gravitational constant, and $F$ is the gas mass ratio
to dynamical mass of a galaxy.  The typical parameters are taken from
the data of BCDs in Table 1. It should be noted that the value of $F$ 
used is a larger than $F$ in Burkert (1995), 0.2, and thus that
our underestimated $\psi_{\rm crit}$ reinforces all gas-rich dwarfs.

Equating both sides of the momentum condition, we obtain
$\psi_{\rm crit}$ once a mass function has been determined. The Salpeter
mass function, $\phi(m)$; $\phi(m)\,\propto\,
m^{-2.35}\,(0.1\,\leq\,m\,\leq\,100\,{\rm M}_{\odot})$ (Salpeter 1955)
is assumed here. Hence,
\begin{eqnarray}
(1+C_{\rm csf}) \psi_{\rm crit}=
1.1 \times \left[\frac{M_{\rm gas}\left(1 + A_{\rm halo} \right)}
{10^8\,M_{\odot}}\right]^{9/7}
\left(1 + A_{\rm halo} \right)^{-2/7}
\,M_{\odot} \,{\rm yr}^{-1},
\label{eq:sal}
\end{eqnarray}
where we adopt the typical values $t_{\rm SF} = 10^7$ yrs, $F = 0.5$,
$M_{\rm gas} = 10^8 ~ M_{\odot}$, $v_{\rm SN}=1000$ km s$^{-1}$, and
$v_{\rm s}=100$ km s$^{-1}$. A correction coefficient of $C_{\rm csf}$
is for the component of continuous star-formation, $C_{\rm csf}=t_{\rm
age}\psi_{\rm cont}/t_{\rm SF}\psi_{\rm crit}$, while $\psi_{\rm cont}$
is a continuous SFR and $t_{\rm age}$ is  galaxy age. Here $C_{\rm
csf}$ is generally less than 1/10 the size of the component of current
star-formation (Legrand et al.  2001; their table 2). Thus, we neglect
it since we are interested in the current condition of the H{\sc i}
envelope.  As long as the observed SFR is well below $\psi_{\rm crit}$,
the H{\sc i} envelope is not blown away by stellar
feedback. If we adopt another IMF and another set of mass ranges, there
is generally 20 percent uncertainty (e.g., Inoue et al. 2000).  

In the next section, we compare $\psi_{\rm crit}$ with the SFR of
dwarfs that is estimated from observations.  For this purpose, we
deduce $\psi_{\rm crit}$ with luminosity of H$\alpha$, $L_{\rm H
\alpha}$ (e.g. Kennicutt 1983), using the relation
$\psi\,=\,2.32\times 10^{-42}\, L_{{\rm H}\alpha} \,M_{\odot}\,{\rm
yr}^{-1} $ that is compiled in and used by Legrand et al. (2001) and
based on Leitherer \& Heckman (1995).  Thus, we can obtain a critical
$L_{\rm H\alpha}$ from eq.(\ref{eq:sal}):
\begin{eqnarray}
L_{\rm H \alpha, \, crit}= 4.1 \times10^{41} \left[\frac{M_{\rm gas}}
{10^8M_{\odot}}\right]^{9/7} \left(1 + A_{\rm halo} \right)^{9/7} {\rm
erg}.
\label{eq:SFR}
\end{eqnarray}

\section{Observational investigation}

\subsection{H{\sc i} envelope and star-formation rate}

We compare the observational data of $L_{\rm H \alpha}$ with
eq.(\ref{eq:SFR}).  The sample galaxies are selected from
\cite{Zee1998}, \cite{Sage1992}, \cite{Zee2000}, and \cite{Zee2001},
according to the condition that they are isolated systems. Regarding
four galaxies found in the several papers, the latest data of each
galaxy are adopted.  We also plot dIrrs in the primary sample of
\cite{Zee2000} and van Zee (2001) to present how well our criterion is
satisfied by dIrrs as well as BCDs.  The sample BCDs are summarized in
Table 1.  The result is depicted in Figure 1. Filled circles are BCDs
and the pluses are dIrrs.  We draw four model lines. The main line is
the thick solid line from eq.(\ref{eq:SFR}) with $A_{\rm halo} =
0.0$. Clearly, all the sample BCDs are located below this line. That
is, all the H{\sc i} envelopes are not blown away by the current
stellar feedback. This result is the same as \cite{Ferrara2000}, since
the dynamical mass-range of our sample galaxies is from $\sim 10^7
M_\odot$ to $\sim 10^9 M_\odot$.  The thick dashed line is from
eq. (\ref{eq:SFR}) with average internal extinction of gas-rich dwarf
galaxies (E(B-V)$=$0.2) (Hunter \& Gallagher 1986) and $A_{\rm halo} =
0.0$.  Even if we
consider the effect of extinction, the conclusion is not altered. The
thin lines are the cases with $A_{\rm halo} = 1.0$. The thin lines are
also sufficiently above the sample BCDs. This strongly indicates that
the H{\sc i} envelope is not blown out by the support of the low
density halo as predicted in Silich \& Tenorio-Tagle (1998).

We find one interesting sample galaxy, Mrk 67, which is plotted at
($M$(H{\sc i}),$L$(H$\alpha$)) $=$ ($10.0^{7.4} M_\odot$,
$10.0^{39.9}$) in Figure 1. It is a Wolf-Rayet galaxy (Conti 1991).
This galaxy is located near the solid line. The H{\sc i} envelope is
strongly affected by stellar feedback. Thus, if we want to know the
precise effect of the low density halo, the detailed observation of
the more extended H{\sc i} envelope of Mrk 67 is important.

\begin{table}
\begin{center}
\begin{tabular}{lcccrcc}
\hline
galaxy&logM(HI)[$M_{\odot}$]&logL(H$\alpha$)[erg]&$F$&12+log(O/H)&[Fe/O]&Ref\\
\hline
IIZw40  & 8.64 & 41.23&0.68&8.13&&a,d\\
UGC4483 & 7.57 & 38.64&0.32&7.52&&a\\
UM439   & 8.49 & 40.07&0.36&8.05&&a\\
UM461   & 8.23 & 40.00&    &7.74&&a\\
UM462   & 8.42 & 40.52&0.51&7.98&&a\\
Haro2   & 8.68 & 41.00&0.31&8.4 &&b,d\\
Haro3   & 8.76 & 41.03&0.47&8.3 &&b,d\\
UM465   & 7.71 & 39.90&    &8.9 &&b\\
Mrk67   & 7.36 & 39.95&0.20&8.09&&b,d\\
Mrk900  & 8.18 & 40.37&    &8.5 &&b\\
Mrk328  & 8.23 & 40.51&0.09&8.5 &&b,d\\
UGC11755& 8.20 & 39.99&    &    &&e\\
UGCA439 & 8.52 & 39.87&0.66&    &&d,e\\
Mrk600  &  8.52 &&  1.51 &  7.83 &  0.17 &c,d\\ 
Mrk5    &  8.11 &&  0.50 &  8.04 &  0.14 &c,d\\ 
Mrk71   &  9.00 &&  0.68 &  7.85 &  0.49 &c,d\\  
IZW18   &  7.81 &&  2.09 &  7.18 &  0.38 &c,d\\  
Mrk22   &  8.18 &&  0.69 &  8.00 &  0.57 &c,d\\  
Mrk36   &  7.28 &&  0.79 &  7.81 &  0.37 &c,d\\  
VIIZw403&  7.58 &&  0.61 &  7.69 &       &c,d\\  
Mrk750  &  7.23 &&  0.18 &  8.11 &  0.50 &c,d\\  
Mrk209  &  7.70 &&  1.13 &  7.77 &  0.49 &c,d\\  
Mrk59   &  8.66 &&  0.36 &  7.99 &  0.50 &c,d\\  
\hline
\end{tabular}
\caption{BCD sample galaxies. The date are taken from a (van Zee et
al. 1998), b (Sage et al. 1992), c (Izotov \& Thuan 1999), d (Thuan \&
Martin 1981) and e (Van Zee 2000) and (van Zee 2001). Blanks mean no
data.}
\end{center}
\end{table}

\subsection{Star-formation history of BCD}

In this paper, we are interested in the star-formation history of
isolated BCDs, which can be given by their chemical abundance in their
star-forming region.  We present this in Figure 2, while the data of
sample galaxies is listed in Table 1.  In panel (a), we plot the
metallicity of BCDs and the gas mass to the dynamical mass ratio. The
metallicity is determined in the current star-forming region (van Zee
et al. 1998; Izotov \& Thuan 1999).  The figure shows that the mass
ratio becomes smaller as the metallicity grows larger. Thus, panel (a)
supports a closed box assumption for the isolated BCDs, as is
originally found in Lequeux et al. (1979). For a detailed review, see
Kunth \& \"Ostlin (2000). Furthermore, as long as a closed box
assumption is satisfied, it can suggest that the H{\sc i} gas in the
envelope is used for the subsequent star-formation in the current
star-forming region. This is consistent with Figure 1, as long as
H{\sc i} gas is generally bound to the host galaxies.  It should be
noted that higher mass ratios than 1 are obtained simply from
observational uncertainties of H{\sc i} gas and/or estimation of
dynamical mass (Thuan \& Martin 1981).  Panel (b) represents the metal
abundance and the same mass ratio. This is also interesting since
there are no significant differences in abundance among the BCDs (The
abundance is located around 0.4, which is a typical value of SN {\sc
ii}).  Panel (b) suggests that the star-formation histories of BCDs
are not very different from each other, as discussed in Kinman \&
Davidson (1981). Thus, we can conclude that almost all BCDs have
experienced a similar star-formation history.

\section{Discussion}

Our results can provide some insight into a morphological
evolutionary model among dwarf galaxies (Kormendy 1985).  The most
standard model for this kind of research, which discusses the H{\sc i}
envelope, is Thuan (1985). They predict that gas-rich dwarfs evolve
into dEs after the H{\sc i} envelope is stripped for some reasons,
such as: (1) interaction with clusters and/or galaxies; and (2)
stellar feedback.  Since the gas-rich dwarf sample in this paper is
isolated, only the stellar feedback effect is expected to be important
for the morphological exchange.  Importantly, the current stellar
feedback does not blow away the H{\sc i} envelope as shown in Figure
1.  Presumably, it does not blow out the H{\sc i} envelope because of
the effect of the low density halo (see also Legrand et al. 2001).
Thus, we conclude that the isolated gas-rich dwarfs can keep their
morphological type, until they no longer experience any
interactions. As long as the the H{\sc i} envelope has been retained,
the kinematics of H{\sc i} envelope is expected to be different from
that of stars in dEs.  This is checked by estimating the specific
angular momentum, and indeed a difference exists (van Zee et
al. 2001).
Thus, our conclusion is consistent with other studies, since isolated
BCDs never evolve into dEs.

In Figure 2, we suggest that the H{\sc i} gas is used for the
subsequent star-formation. This might permit BCDs and dIrrs to evolve
to dEs.  However, the consumption time-scale, which is defined as the
mass of the H{\sc i} envelope divided by the current star-formation
rate, should be longer than $10^{10}$ years. It is difficult for
current BCDs and dIrrs to evolve into dEs.  Our discussion may confirm
the observational implication on its difficulty (Ferguson \& Binggeli
1994) in terms of a SFR.  We comment that the photoionization effect
can be important during the evolution from dIrrs to dEs (Ferrara \&
Tolstoy 2000), while this effect cannot be examined in our
framework. This possibility, related to photoionization effects, will
be discussed in a future work.

There is another possibility for the morphological change between
dIrrs and BCDs. According to Sait\=o et al. (2000), gas-rich dwarfs
change their morphology between BCDs and dIrrs.  This means that the
star-formation activity is related to the extent of the H{\sc i}
envelope.  Here, we re-consider this evolution in Figure 1: dIrrs with
massive H{\sc i} gas move upper-leftward in Figure 1 and are
recognized to be BCDs, when the H{\sc i} gas falls onto galaxies and
is consumed for the successive star-formation period. On the contrary,
when BCDs decrease their star-formation, they move downward in Figure
1, and are recognized as dIrrs.  Thus, our plot of Figure 1 is
consistent with the preliminary idea of Sait\=o et al.(2000). The
trend of consumption of H{\sc i} mass in Figure 2(a) may support this
evolutionary model of gas-rich dwarfs.

An inverse evolutionary scenario from dEs to gas-rich dwarfs has been
suggested by Silk et al. (1987).  The IGM was heated and enriched by
galactic winds at higher redshift. When galaxy groups formed recently,
this gas was compressed, cooled and accreted onto dwarf galaxies with
some metallicity. Thus, dEs evolve into dIrrs and BCDs.  This scenario
is re-interpreted if we exchange the IGM for the low density halo
around the BCDs. Since the low density halo is expected to be bound to
the galaxy, a closed box assumption can also be invoked
here. Furthermore, both IGM and the low density halo have a similar
role in the dynamical evolution of H{\sc i} envelope (Silich \&
Tenorio-Tagle 1998). Also, if a past SFR was larger than $\psi_{\rm
crit}$, the H{\sc i} envelope could have been blown away from the host
dwarf galaxy, while being retained in the galaxy by the low density
halo. Hence, when future observations will reveal the large past SFR
of BCDs, unlike current observations of their spectral energy
distribution, then, it will be interesting for us to re-examine the
possibility of the morphological evolution model of Silk et al. (1987)
with caution.

\section{ Summary }

In this paper, we have shown that the H{\sc i} envelope are not easily
blown away by stellar feedback without the low density halo but with
dark matter, since the envelope is not accelerated to the escape
velocity (see also Legrand et al. 2001). This dynamical state is
observationally confirmed for isolated gas-rich dwarfs.  Furthermore,
we have confirmed the possibility that all the isolated BCD samples
have a similar star-formation history, as is suggested from their
metallicity and H{\sc i} mass consumption.  Our results support that
(1) It is difficult for the isolated gas-rich dwarfs to evolve into
dEs, since the present SFRs of gas-rich dwarfs are not so large.  They
will change their morphological types only when they suffer
exceptionally intense star-burst and/or interaction among galaxies,
and/or clusters of galaxies.  (2) The morphological exchange among
gas-rich dwarfs may be possible. But, as a future work, we must
describe the star-formation mechanism due to the H{\sc i} gas fueled
into star-forming regions, suggested from the anti-correlation between
H{\sc i} mass and metallicity.  The effect of the low density halo may
be essential.  (3) A morphological evolutionary model from dEs to
gas-rich dwarfs is possible if the H{\sc i} gas falls onto the host
galaxy. But, even in this case, the low density halo should be
investigated further. Recently, Legrand et al. (2001) has reported a
very similar result to our Figure 1. The essential meaning of the two
results are the same and consistent with each other, such that the ISM
is sustained in the BCDs themselves. However, we consider especially
the H{\sc i} envelope, and find that the H{\sc i} envelope is neither
blown away nor blown out from BCDs by means of the current stellar
feedback.

\acknowledgements

Critical comments of the referee, Dr. Daniel Kunth, much improved the
clarity of the paper and our English.  The authors appreciate his very
careful reading and much kindness. They are also grateful to
H. Hirashita for valuable comments and discussion. HK wishes to thank
Prof. S.Mineshige, S.Inagaki, and J.Silk for their encouragement.



\begin{figure}
\includegraphics{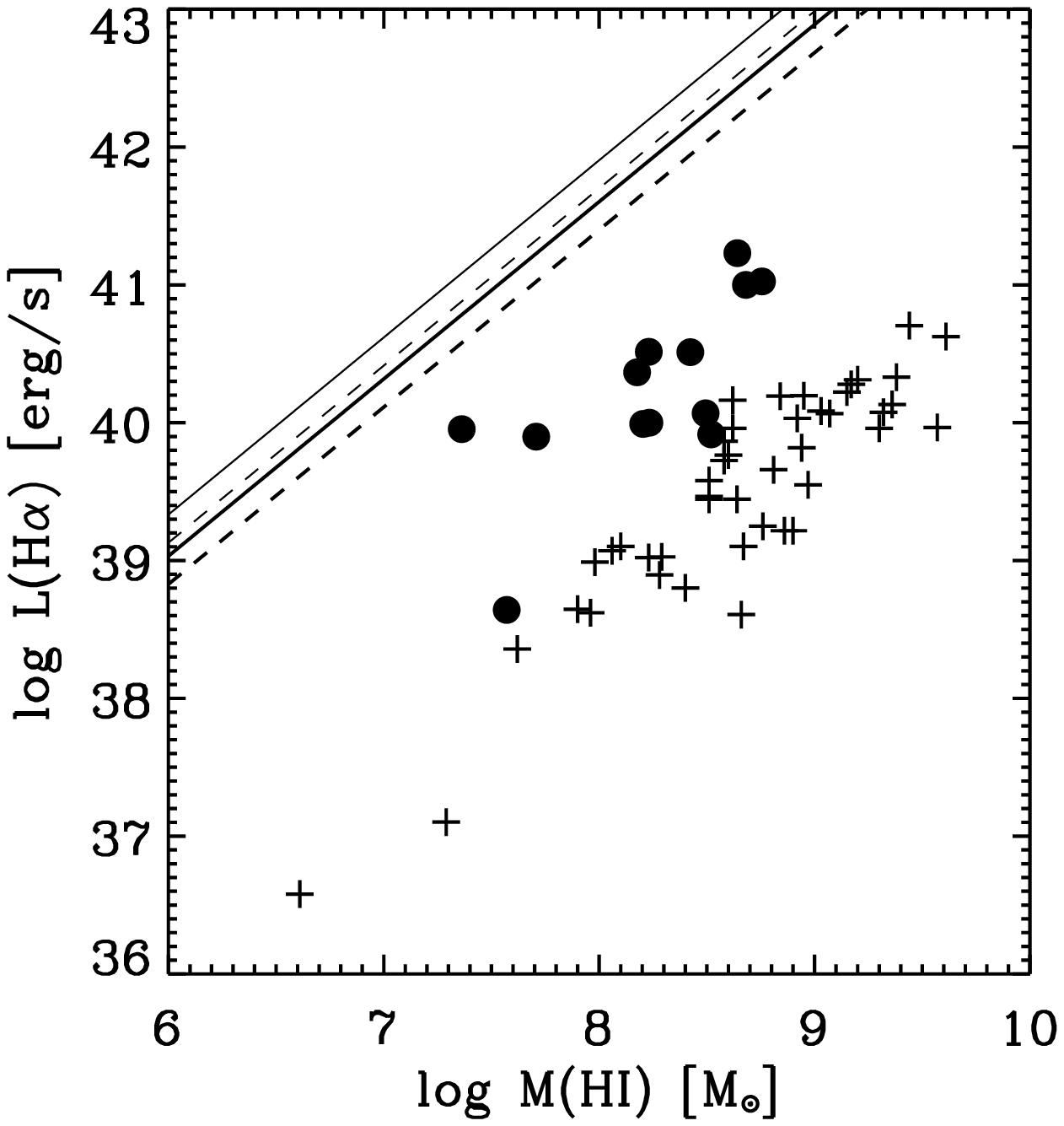} \caption{The luminosity of H$\alpha$
and the mass of H{\sc i} gas instead of $M_{\rm gas}$.  The filled
circles are BCDs and the pluses are dIrrs.  The solid lines are from
eq. (\ref{eq:SFR}) and the dashed lines are from eq. (\ref{eq:SFR})
with average internal extinction in dwarf galaxies (E(B-V)=0.2). The
thick and thin lines are from eq. (\ref{eq:SFR}) at $A_{\rm halo}=0$
and $A_{\rm halo}=1$, respectively.}  \label{HI_SFR}
\end{figure}

\begin{figure}
\includegraphics{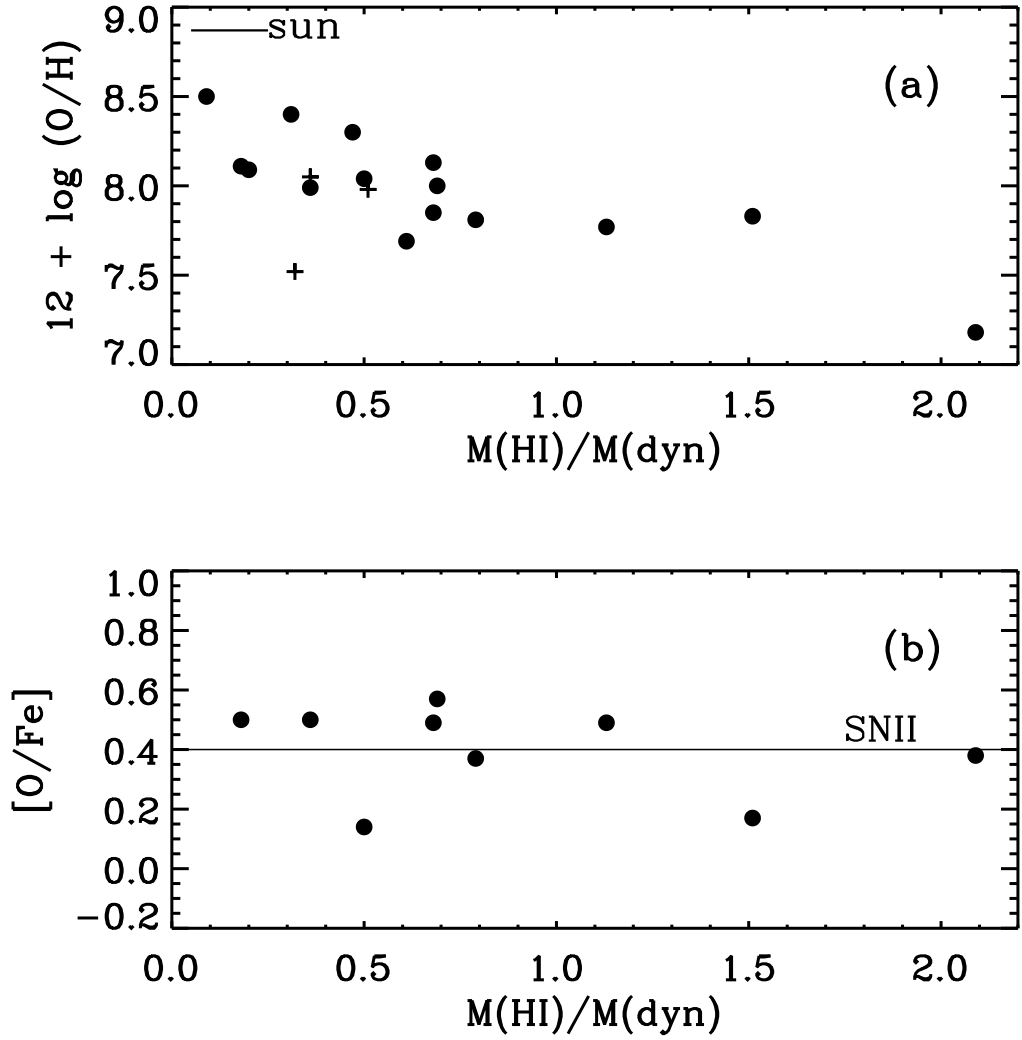} \caption{\textbf{a)} The
anti-correlation between the metallicity and the ratio of H{\sc i} gas
mass to the dynamical mass of samples in Table 1. The filled circles
denotes samples with the mass ratio from Thuan \& Martin (1981), and
the plus denotes those of \cite{Zee1998}.  \textbf{b)} The flat
distribution of metallicity abundance to H{\sc i} gas mass ratio.  The
metal enrichment is determined at the current star-forming region. }
\label{HI_Z}
\end{figure}

\begin{thebibliography}{}
 \bibitem[Barone et al. (2000)]{Barone2000} Barone, L.T., Heithausen,
			       S., H\"{u}ttemerise, S., Fritz, T., \&
			       Klein, U. 2000, MNRAS, 317, 649
 \bibitem[Bothun et al. (1986)]{betal86} Bothun, G.D., Mould, J.R.,
			       Caldwell, N., \& Mac Gillivray,
			       H.T. 1986, AJ, 92, 1007
 \bibitem[Burkert (1995)]{Burkert1995} Burkert, A. 1995, ApJL, 447, 25
 \bibitem[Chevalier (1976)]{Chivalier1976} Chevalier, R.A. 1976,
                              ApJ, 207, 872
 \bibitem[Conti (1991)]{conti91} Conti, P.S. 1991, ApJ, 377, 115
 \bibitem[Dekel \& Silk (1986)]{Dekel1986} Dekel, A, \& Silk,
                              J. 1986, ApJ, 303, 39
 \bibitem[De Young \& Heckman (1994)]{YH94} De Young, D.S., \& Heckman,
			       T.M. 1994, ApJ, 431, 598
 \bibitem[Ferguson \& Binggeli (1994)]{FB94} Ferguson, J.W., \&
                              Binggeli, B., 1994, ARA\&A, 6, 67
 \bibitem[Ferrara \& Tolstoy (2000)]{Ferrara2000} Ferrara, A., \&
			       Tolstoy, E. 2000, MNRAS, 313, 291
 \bibitem[Gil et al. (1999)]{Gil1999} Gil de Paz, A., Zamorano, J., \&
			       Gallego, J. 1999, MNRAS, 306, 975
 \bibitem[Hunter \& Gallagher (1986)]{Hunter1986} Hunter, D.A., \&
			       Gallagher, J.S. 1986, PASP, 98, 5
 \bibitem[Inoue et al. (2000)]{Inoue2000} Inoue, A.K., Hirashita, H., \&
			       Kamaya, H. 2000, PASJ, 52, 539
 \bibitem[Izotov \& Thuan (1999)] {Izotov1999} Izotov, Y., \& Thuan,
			       T.X.  1999, ApJ, 511, 639
 \bibitem[Jones et al. (1981)]{Jones1981} Jones, E.M., Smith, B.W., \&
			       Straka, W.C. 1981, ApJ, 249, 185
 \bibitem[Kennicutt (1983)]{Kennicutt1983} Kennicutt, R.C. 1983, ApJ,
			       272, 54
 \bibitem[Kinman \& Davidson (1981)]{kd81} Kinman, T.D., \& Davidson,
			       K. 1981, ApJ, 243, 127
 \bibitem[Kormendy (1985)]{K85} Kormendy, J. 1985, ApJ, 295, 73 
 \bibitem[Kunth \& \"Ostlin (2000)]{KO00} Kunth, D., \& \"Ostlin,
			       G. 2000, ARA\&A, 10, 1
 \bibitem[Lamers \& Cassinelli(1999)]{Lamers1999} Lamers, H.G.L.M., \&
			       Cassinelli, J.P. 1999, in {\it
			       Introduction to Stellar Winds}, \S12.6
 \bibitem[Larson (1974)]{l74} Larson, R.B. 1974, MNRAS, 169, 229 
 \bibitem[Legrand et al. (2001)]{Letal01} Legrand, F., Tenorio-Tagle,
			       G., Silich, S., Kunth, D., \& Cervi\~no,
			       M. 2001, ApJ, 560, 630
 \bibitem[Lequeux et al. (1979)]{Letal79} Lequeux, J., Peimbert, M.,
			       Rayo, J.F., Serrano, A., \&
			       Torres-Peimbert, S. 1979, A\&A, 80, 155
 \bibitem[Martin (1996)]{Martin1996} Martin, L.C. 1996, ApJ, 465, 680
 \bibitem[Mac Low \& Ferrara (1999)]{MF99} Mac Low, M.-M., \& Ferrara,
			       A.  1999, ApJ, 513, 142
 \bibitem[Pustilnik et al. (2001)]{Pustlinik01} Pustilnik S.A., Brinks,
			       E., Thuan, T.X., Lipovetsky, V.A., \&
			       Izotov, Y.I. 2001, AJ, 121, 1413
 \bibitem[Sage et al. (1992)]{Sage1992} Sage, L. J., Salzer,
			       J. J., Loose, H.H., \& Henkel, C. 1992,
			       A\&A, 265, 19
 \bibitem[Salpeter (1955)]{Salpeter1955} Salpeter, E.E. 1955, ApJ, 121,
			       161
 \bibitem[Saito (1979)]{Saito1979} Saito, M. 1979, PASJ, 31, 193
 \bibitem[Sait\=o et al. (2000)]{skt01} Sait\=o, M., Kamaya, H., \&
			       Tomita, A. 2000, ASP Conf. Ser.218, 305
 \bibitem[Silk et al. (1987)]{ses87} Silk, J., Wyse, F.G.,
			       \& Shields, G.A. 1987, ApJ, 322, L59
 \bibitem[Skillman \& Bender (1995)]{sb95} Skillman, E.D., \& Bender,
			       R. 1995, Rev.Mex.AA, 3, 25
 \bibitem[Silich \& Tenorio-Tagle (1998)]{ST98} Silich, A.S., \&
			       Tenorio-Tagle, G. 1998, MNRAS, 299, 249
 \bibitem[Silich \& Tenorio-Tagle (2001)]{ST01} Silich, A.S., \&
			       Tenorio-Tagle, G. 2001, ApJ, 552, 91
 \bibitem[Silich et al. (2001)]{Setal01} Silich, A. S., Tenorio-Tagle,
			       G., Terlevich, R., Terlevich, E., \&
			       Netzzer, H., 2001, MNRAS, 324, 191
 \bibitem[Thuan (1985)]{Thuan1985} Thuan, T.X. 1985, ApJ, 299, 881
 \bibitem[Thuan \& Martin (1981)]{Thuan1981} Thuan, T.X., \& Martin,
			       G.E. 1981, ApJ, 247, 823
 \bibitem[van Zee et al. (1998)]{Zee1998} van Zee, L. et al. 1998, AJ,
			       116, 1186
 \bibitem[van Zee (2000)]{Zee2000} van Zee, L. 2000, AJ, 119, 2757
 \bibitem[van Zee (2001)]{Zee2001} van Zee, L. 2001, AJ, 121, 2003
 \bibitem[van Zee et al. (2001)]{Zee.et.al2001} van Zee, L., Salzer, J.J., \&
			       Skillman, E.D. 2001, AJ, 122, 121
\end{thebibliography}
\end{document}